\documentclass[aps,pre,twocolumn,showpacs,floatfix,amssymb,preprintnumbers]{revtex4}

\usepackage[dvips]{graphicx}
\usepackage{dcolumn}
\usepackage{bm}
\usepackage[dvips]{epsfig}
\usepackage{amssymb}
\usepackage{latexsym}

\begin{document}
 
\newcommand \be {\begin{equation}}
\newcommand \ee {\end{equation}}
\newcommand \bea {\begin{eqnarray}}
\newcommand \eea {\end{eqnarray}}

\title{Intermittent outgassing through a non-Newtonian fluid}

\author{Thibaut Divoux, Eric Bertin, Val\'erie Vidal and Jean-Christophe G\'eminard}
\affiliation{Universit\'e de Lyon, Laboratoire de Physique, Ecole Normale Sup\'erieure de
Lyon, CNRS, 46 All\'ee d'Italie, 69364 Lyon cedex 07, France.}

\begin{abstract}

We report an experimental study of the intermittent dynamics of a gas flowing through a column of a non-Newtonian fluid.
In a given range of the imposed constant flow rate, the system spontaneously alternates between two regimes: bubbles emitted at the bottom either rise independently one from the other or merge to create a winding flue which then connects the bottom air-entrance to the free surface. The observations are reminiscent of the spontaneous changes in the degassing regime observed on volcanoes and suggest that, in the nature, such a phenomenon is likely to be governed by the non-Newtonian properties of the magma. We focus on the statistical distribution of the lifespans of the bubbling and flue regimes in the intermittent steady-state. The bubbling regime exhibits a characteristic time whereas, interestingly, the flue lifespan displays a decaying power-law distribution. The associated exponent, which is significantly smaller than the value $1.5$ often reported experimentally and predicted in some standard intermittency scenarios, depends on the fluid properties and can be interpreted as the ratio of two characteristic times of the system. 

\end{abstract}


\maketitle

\section{Introduction}

A large number of physical systems driven far from equilibrium exhibit an intermittent behavior, where quiet phases (\textsl{off}-state) randomly alternate with periods of strong activity (\textsl{on}-state). Experimental examples in physical systems include electronic devices \cite{Hammer94}, spin-wave instabilities \cite{Rodelsperger95}, plasma produced by a gas discharge \cite{Feng98}, liquid crystals \cite{John99,Vella03}, and nanoscopic systems such as single molecules \cite{Dickson97}, nanocrystal quantum-dots \cite{Brokmann03}, or nanoparticles diffusing through a laser focus \cite{Zumofen04}. Intermittency has also been reported in other disciplinary fields, like in the example of the sleep-wake transitions during sleep \cite{Lo2002}.
A standard characterization of the intermittency phenomenon is the distribution of time spent in the {\it off}-state.
Interestingly, in most of the experimental studies reported so-far \cite{Hammer94,Rodelsperger95,Feng98,John99,Vella03,Dickson97,Brokmann03,Zumofen04}, this distribution exhibits a power-law regime with an exponent close to $1.5$, with a cut-off at large times. This value of the exponent is well-understood theoretically through the so-called scenarios of {\it on-off} intermittency \cite{Fujisaka1985,Platt1993,Heagy1994} and type-III intermittency \cite{Pomeau1980,Schuster1989},
which both predict a power-law distribution with an exponent $3/2$.

A Newtonian fluid flowing through a non-Newtonian material is alike to present different escaping regimes \cite{Zoueshtiagh} as well as an intermittent activity, as previously observed both in physical \cite{Gostiaux,Varas} and geophysical \cite{Ripepe,Gonnermann} contexts. In the laboratory, experiments have been conducted in the case of air emission through an immersed granular layer \cite{Gostiaux}. These studies evidenced different air-outgassing regimes through the non-linear medium at a constant flow rate. In particular, it has been observed that gas emitted at the bottom of an immersed granular layer can cross the whole system through temporary channels, which later spontaneously collapse \cite{Gostiaux,Varas}.
In a geophysical context, volcanologists reported observations of the degassing process which interestingly exhibits similar temporal patterns \cite{Ripepe}. In this case, the gas, released by magma during its rise toward the free surface, plays the role of the Newtonian fluid and magma the role of the non-Newtonian material \cite{Webb90,Lavallee2007}: inside the volcano conduit, dissolved gas gives birth to bubbles, which ascend and burst intermittently at top of the volcano vent \cite{Gonnermann}.
What remains puzzling is the existence of spontaneous changes between different activities ({\it e.g.} discrete bursts or continuous puffing \cite{Parfitt}) for which the non-Newtonian properties of magma could play an important role \cite{Gonnermann,Lavallee2008,Caricchi08}.

At the laboratory scale, two kinds of experiments have been previously conducted in order to model the volcanic degassing-activity. On the one hand, experiments were designed to account for the potential effects of the geometry of an underlying magma-chamber \cite{Jaupart}. These studies were limited to the case of Newtonian fluids and, besides, did not consider the issue of the time spent in each of the regimes.
On the other hand, recent laboratory experiments performed on real rhyolitic melts close to their natural conditions have revealed that during their rise, bubbles deform and coalesce under shear \cite{Okumura2006} leading to channel-like bubble networks \cite{Okumura2008}. Those structures were proposed as a key ingredient in the magma degassing process at depth \cite{Okumura2008}, but not directly related to the non-Newtonian rheology of magma.

In this paper, we report an experimental study of the intermittent dynamics of a gas rising through a column of a non-Newtonian fluid. We unravel that this outgassing process can be considered as a two-states process : either it occurs through bubbles rising independently  ({\it the bubbling regime}, {\it on}-state), or through a gas channel which crosses tortuously the whole system ({\it the winding-flue regime}, {\it off}-state). Let us emphasize here that such an intermittent process, defined in terms of a two-states process (bubbling or channel-like regime), is different from the intermittent regime reported in purely Newtonian fluids, as described in \cite{Tufaile}. Indeed, a Newtonian solution cannot sustain the formation of a flue
(in contrast to non-Newtonian fluids \cite{Kliakhandler}) and, in this case, intermittency can be related to the formation of antibubbles \cite{Tufaile}. Here, we report statistical data concerning the irregular oscillations between the bubbling and the winding flue regimes.
Specifically, we measure the statistics of the time spent in both the bubbling and flue regimes.
We observe that the lifespan of the flue is characterized by a power-law distribution with an exponent significantly smaller than the often reported exponent $3/2$. This non-standard exponent turns out to depend slightly on the rheological properties of the fluid. Using a simple stochastic two-states model, we relate this exponent to the ratio of two characteristic times of the system, that we estimate from the experimental data.

These results might be of interest to the geophysics community as the magma, charged with bubbles, exhibits a flow-threshold and a shear-thinning behavior \cite{Webb90,Gonnermann,Lavallee2007,Pinkerton95}. Indeed, non-Newtonian rheological properties are likely to explain the existence of different outgassing processes (including bubble networks) and to govern the associated nontrivial statistics. 

\section{Experimental procedures}

\subsection{Experimental setup}

\begin{figure}[t]
\begin{center}
\includegraphics[width=0.9\columnwidth]{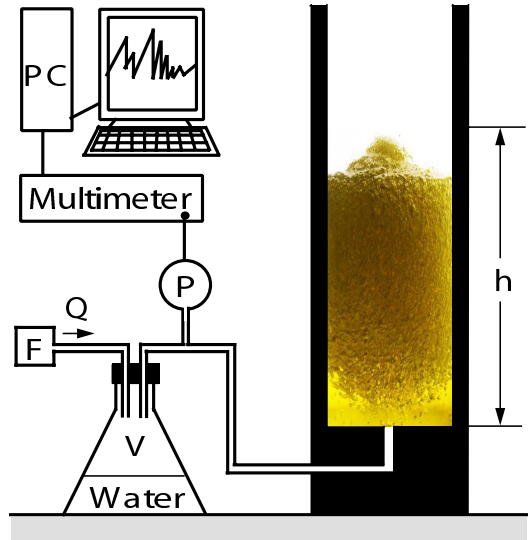}
\end{center}
\caption{\small{\bf Sketch of the experimental setup -}
Air, from a mass-flow controller $\fbox{F}$, is injected, at a constant flow rate $Q$, in a chamber (volume $V$) connected to the bottom of the fluid column (height $h$). By means of a presure sensor $P$, we measure the overpressure $\delta P$ inside the chamber as a function of time. In the picture on the right-hand-side, one can notice that the density of bubbles trapped into the gel strongly depends on the altitude (The scale is given by the inner diameter of the plexiglass tube, $74$~mm.)}
\label{fig.schema}
\end{figure}

The experimental setup (Fig.~\ref{fig.schema}) consists of a vertical plexiglass tube (inner diameter $74$~mm, height $270$~mm) filled with a non-Newtonian solution.
Air is injected at the bottom of the fluid column through an air tank, partially filled with water. The system thus makes it possible to tune the inner free volume $V$ of the tank (ranging from $60$ to $800$~cm$^3$) by changing the inner amount of water and to inject humid air. Besides, above the fluid column, a small container filled with water maintains a saturated humidity level at the free surface. This way, one avoids any significant drying of the sample during the experimental time. A mass-flow controller (Bronkhorst, Mass-Stream series D-5111) introduces air inside the tank at a constant flow rate $Q$ (from $Q_{min}=0.17$~cm$^3$/s to $Q_{max}=1.74$~cm$^3$/s). Air is then injected at the bottom of the fluid column through a nozzle (diameter $d=2.0$~mm). Rigid tubes (typical diameter, $8.0$~mm) insure the connection between the tank and the column. We measure the variation of the overpressure $\delta P$ inside the chamber with the help of a low-pressure differential-sensor (Honeywell S$\&$C, 176PC28HD2) connected to a multimeter (Keithley, 196). A C++ program records the pressure data (to within 5 Pa) through the GPIB interface over long time-durations (typically a few days) at 5~Hz. As we will see, monitoring the pressure in the tank chamber makes it possible to probe the degassing activity while direct visual observation is obstructed as the gel solution progressively fills up with bubbles (Sec.~\ref{procedure}).

The fluid consists in a diluted solution ($10\%$ in mass of distilled water, unless otherwise specified) of a commercial hair-dressing gel (\textsl{Gel coiffant, fixation extra forte},  Auchan). This choice is mainly justified by the fact that one can easily be supplied with large quantities of fluid, reproducible mixtures are rather easy to prepare \cite{Divoux} and they are stable in time if well prevented from drying. Some of the non-Newtonian rheological properties of the fluid are reported in the appendix A. In particular, the solution, when free of bubbles, is shear-thinning and we estimated the viscoelastic characteristic time $\tau_c = 1/\dot \gamma _c \simeq 1700$~s, compatible with the ratio of the viscosity at low shear-rate $\eta_0 \simeq 6.10^4$~Pa.s to the estimated flow threshold $\sigma _c \simeq 35$~Pa.

\subsection{Preliminary observations}
\label{po}

The gel solution
is poured inside the tube and a constant flow is imposed. Qualitatively, one observes that the solution starts filling up with small-sized bubbles.
These small bubbles, which remain trapped because of the fluid threshold, are further advected by the large bubbles subsequently emitted at the nozzle and progressively a vertical gradient pointing toward the top of the column takes place (Fig.~\ref{fig.schema}). After this transient regime, whose duration depends on the flow rate $Q$, the fluid column reaches a constant height $h$, significantly larger than the initial height $h_0$. We observe that there exists a flow rate $Q_b$ which can be experimentally defined as the most efficient flow rate to fill the solution with trapped bubbles. The transient which lasts a few hours for $Q \simeq Q_b$ can last days to weeks if $Q$ significantly differs from $Q_b$. Special attention has been paid to the characterization of the bubble gradient and of the associated fluid rheology. Details are reported in the appendix A. Finally, note that all the data reported in this article were obtained after the bubble gradient was established.

\subsection{Experimental protocol}
\label{procedure}

We pointed out that small bubbles remain trapped in the fluid, which might lead to an evolution of the rheological properties through time (aging). Thus, in order to ensure the reproducibility of the experiment, we chose the following experimental protocol : the gel solution, initially free of bubbles, is poured inside the tube; the flow rate is set to the smallest accessible value, $Q_{min}$; the overpressure $\delta P$ is then recorded for, at least, 3 days. Subsequently, the flow rate is increased by steps of amplitude $\delta Q$ up to the maximum value, $Q_{max}$. For each value of the flow rate $Q$, the pressure $P$ is again recorded for, at least, 3 days. Then, the same process is applied to decrease the flow rate $Q$ down to $Q_{min}$. The whole process will be referred to as a {\it flow rate cycle}. For all the experiments reported here, we chose $\delta Q \simeq Q_{min}$, corresponding to about ten data points over the experimental range of $Q$.

\begin{figure}[t]
\begin{center}
\includegraphics[width=0.9\columnwidth ]{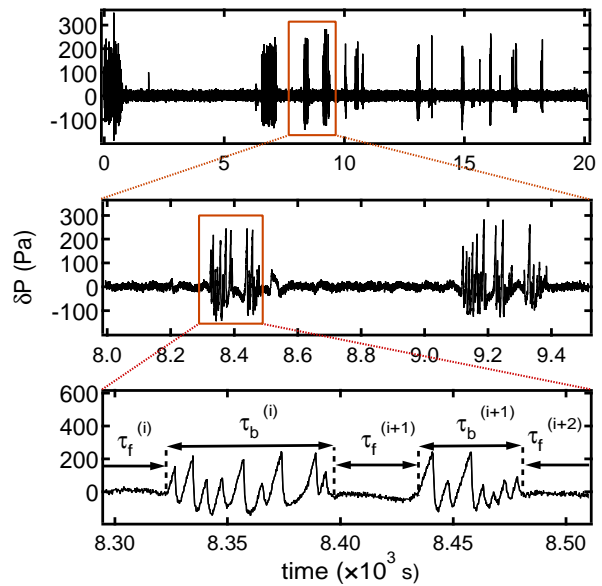}
\end{center}
\caption{\small{{\bf Typical pressure variation $\delta P(t)$ vs time $t$ -} 
At a constant flow rate, the typical signal from the pressure sensor exhibits a spontaneous alternation between rest and activity periods. Rapid drops of the overpressure mark the emission of successive bubbles at the bottom of the column whereas an almost-constant overpressure corresponds to a flue connecting the injection hole to the free surface. The analysis consists of measuring the lifespan $\tau_b$ (resp. $\tau_f$) of the bubble (resp. flue) regime. The figures from top to bottom display the signal at different time scales ($h_0=70$~mm, $V=800$~cm$^3$ and $Q=0.35$~cm$^3$/s).}}
\label{fig.2}
\end{figure}

\section{Experimental results}
\label{Experimental}

\subsection{Overall behavior of the system}

We now turn to the study of the pressure recorded inside the tank-chamber. We report $\delta P$, the variations of the pressure around its mean value, high-pass filtered at $1$~mHz. One immediately notices that the pressure exhibits two distinct behaviors (Fig.~\ref{fig.2}), which emphasizes rest and activity periods.

Thus, a first crucial result is the following: for a given flow rate, the system spontaneously alternates, in an irregular manner, between two states. Either cusped-like bubbles are emitted at the bottom of the gel column, rise through it and burst at the top independently from one another (the \textsl{bubbling} regime); the corresponding pressure variation $\delta P$ exhibits successive rises and drops, the latter corresponding to the formation of the bubbles. Or, a channel, previously reported as a 'bubble sausage' \cite{Kliakhandler}, connects the bottom air-entrance to the column top (\textsl{winding flue} regime); the corresponding $\delta P$ is almost constant. 

The formation of the flue is linked to the non-Newtonian properties of the hair-gel solution and, as we shall see, to the thixotropy. A rising bubble produces a shear which locally leads to a decrease of the fluid resistance to flow, thus creating an easy path through the fluid column \cite{Daugan}. This process takes place if the time-difference between two successive bubbles is shorter than the typical relaxation time scale of the fluid. Once formed, the flue, subjected to a creep flow under the effect of gravity, slowly deforms and eventually collapses, leading the system back to the bubbling regime. These two mechanisms explain why the system switches quickly between two very different states. 

\begin{figure}[t]
\begin{center}
\includegraphics[width=0.9\columnwidth]{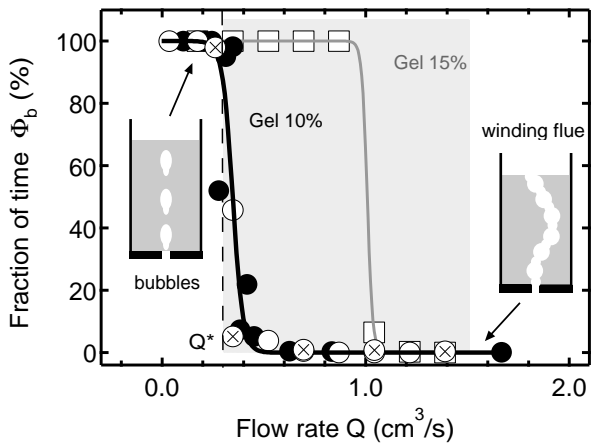}
\end{center}
\caption{\small{\textbf{Fraction $\Phi_b$  vs. imposed flow rate $Q$ -}
The fraction of time spent in the {\it bubbling} regime, $\Phi_b$, clearly exhibits a transition at a given flow rate $Q^*$ which drastically depends on the fluid rheology. 
Fitting curves, sigmoids, are given as eye leads. The gray area marks the range, 
$Q \in\,]Q^*,Q^+]$, in which the duration of the bubbling regime can be defined (associated with $\bullet $, only) [(symbol:~height~$h_0$ mm,~volume~$V$~cm$^3$). Added-water content $10~\%$: ($\bullet$:~$70$,~$530$); ($\circ$:~$130$,~$530$);
($\otimes$:~$70$,~$800$). Added-water content $15~\%$: ($\square$:~$70$,~$530$)].}}
\label{fig.pourcent}
\end{figure}

As a first global result, we report the fraction of time, $\Phi_b$, spent by the system in the \textsl{bubbling} state at a given flow rate (Fig.~\ref{fig.pourcent}). Below a certain threshold $Q^*$, the system experiences the {\it bubbling}-regime, whereas above $Q^*$ the system naturally alternates between the {\it bubbling} and {\it winding-flue} regimes
(We note that $Q^* \simeq Q_b$, which is discussed in the appendix B).
We point out that the two regions, above and below the transition, are not symmetric:
indeed, for $Q < Q^*$, we only observe the bubbling regime (the system is unable to form a flue) whereas, in the case $Q > Q^*$, even for the largest flow rate, the flue always eventually tweaks and collapses (finite lifespan). The critical flow rate $Q^*$ is independent of the column height $h_0$ and of the chamber volume $V$ but depends on the rheological properties of the fluid. For instance, the minimum flow rate necessary to sustain an open flue increases when the fluid threshold is decreased by dilution and we assessed $Q^* \simeq 0.3$ cm$^3$/s for 10\% of added water and $Q^*~\simeq~1$~cm$^3$/s for 15\% (Fig.~\ref{fig.pourcent}). Finally, note that, even if the channel can always collapse at large flow rate, one can identify a flow rate $Q^+ \simeq 1.5$~cm$^3$/s above which the emission of a single bubble leads to the formation of a new channel (flow rate $Q$ above the gray region in Fig.~\ref{fig.pourcent}).

\begin{figure}[t]
\begin{center}
\includegraphics[width=0.9\columnwidth]{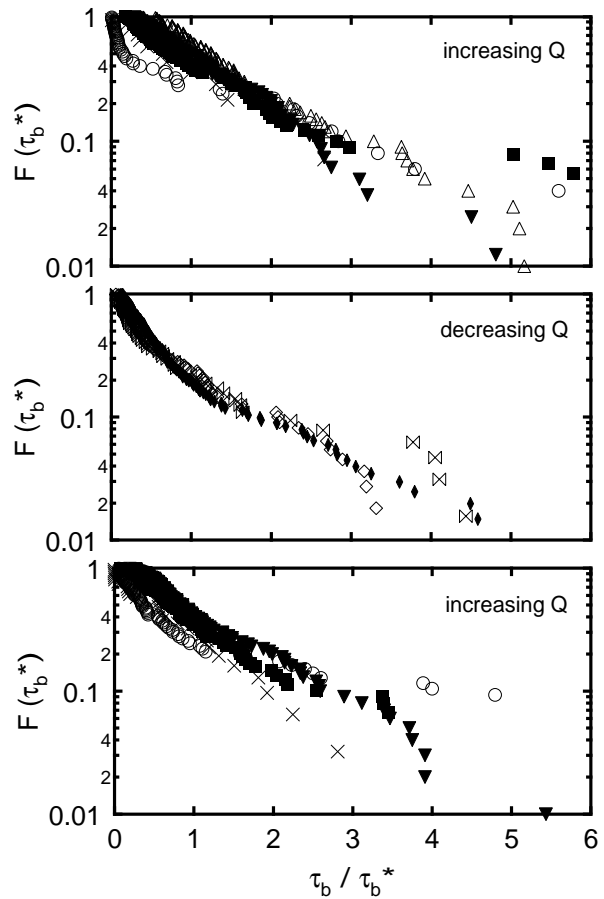}
\end{center}
\caption{\small {\bf Cumulative probability distribution $F_b(\tau_b)$ -} The time $\tau_b$ spent in the bubbling regime exhibits an exponential distribution with a characteristic time $\tau_b^*$ (Fig.~\ref{fig.taubstar}). We report data obtained when increasing and decreasing the flow rate $Q$ [(symbol, $Q$~cm$^3$/s). 
Top: ($\times$, 0.347); ($\circ $,~0.521); ($\blacksquare$,~0.695); ($\triangle$,~0.869);  ($\blacktriangledown $,~1.04);   
Middle: ($\lozenge$,~1.22); ($\blacklozenge$,~0.782); ($\diamond$,~0.608); ($\Join$,~0.434).
Bottom: ($\times$,~0.347); ($\circ$,~0.521); ($\blacksquare$,~0.695); ($\blacktriangledown $,~1.04). (Added-water content $10\%$, $V=530$~cm$^3$, $h_0=70$ mm)].}
\label{fig.distribulle} 
\end{figure}

From now on, because of the binary behavior of gel column at a given flow rate $Q$, we shall account for the complex dynamics of the system by considering the overpressure signal as a temporal sequence
$\{\tau_f^{(1)}, \tau_b^{(1)}, \tau_f^{(2)}, \tau_b^{(2)},..., \tau_f^{(n)}, \tau_b^{(n)}\}$
of the successive {\it flue}- (resp. {\it bubbling}-) regime lifespans, $\tau _{f}$ (resp. $\tau _{b}$).
We aim at estimating the probability distributions $p_b(\tau _{b})$ and $p_f(\tau _{f})$.
To do so, we display estimates of the corresponding cumulative probability
distributions $F_b(\tau_b)$ (Fig.~\ref{fig.distribulle}) and
$F_f(\tau_f)$ (Fig.~\ref{fig.distriCO}), defined as
\be
F_{x}(\tau_{x}) \equiv \int_{\tau_{x}}^{\infty} p_{x}(\tau)\,d\tau
\ee
where $x$ stands for $b$ or $f$.
We sort the sequence $\{\tau_x^{(i)}\}$ into $\{\tau_x'^{(j)}\}$ such that
\be
\tau_x'^{(1)}>\tau_x'^{(2)}>\ldots>\tau_x'^{(n)}
\ee
and plot $j/n$ versus $\tau_x'^{(j)}$ \cite{Berg}.

\subsection{Bubbling regime}

\begin{figure}[t]
\begin{center}
\includegraphics[width=0.9\columnwidth]{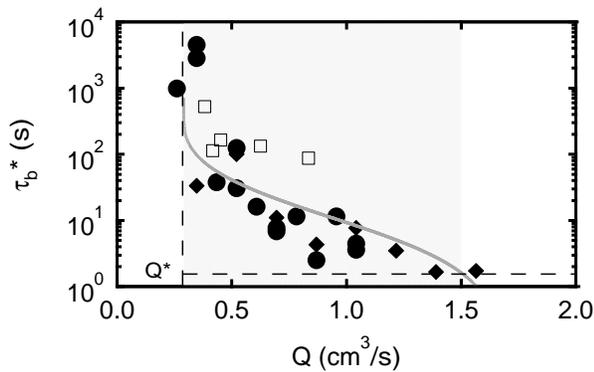}
\end{center}
\caption{\small {\bf Time $\tau_b^*$ vs. flow rate $Q$ -}
Values of $\tau_b^*$ obtained for both increasing and decreasing flow rates are reported. The closed symbols denote the values deduced from the data reported in the figure~\ref{fig.distribulle}. The open symbols correspond to data obtained when the fluid column is stirred after each change in the flow rate ($\square$, $V=530$~cm$^3$). The continuous line corresponds to the equation (\ref{temps}) with $\tau_0 \simeq 65$~s, $Q_c = 0.35$~cm$^3$/s and $Q^* \simeq 0.29$~cm$^3$/s. The horizontal dashed line denotes the characteristic time of emission of a single bubble, about 1~s. The grey region corresponds to $Q \in~] Q^*, Q^+]$ [(Symbol, $V$ cm$^3$) : ($\bullet$, 530); ($\blacklozenge$, $800$). (Added-water content $10\%$, $h_0=70$~mm)].}
\label{fig.taubstar} 
\end{figure}

From the temporal sequence $\{\tau_b^{(n)}\}$, we deduce that there exists a characteristic time, $\tau_b^*$, to build a flue from the bubbling regime. Indeed, the cumulative probability distribution $F_b(\tau_b)$ decreases almost exponentially for increasing values of $\tau_b$ (Fig.~\ref{fig.distribulle}). Thus, the lifespan $\tau_b$ of the {\it bubbling} regime presents an almost-exponential probability distribution. We report the associated characteristic time $\tau_b^*$ as a function of the flow rate $Q$ for different values of the chamber volume $V$ (Fig.~\ref{fig.taubstar}). First, $\tau_b^*$ decreases drastically when the flow rate is increased above $Q^*$ (Note that $\tau_b^*$ is not defined for $Q<Q^*$ as the system is constantly in the bubbling regime and that, for $Q>Q^+$, the measurements are difficult to perform because of the small number of channel collapses during the experimental time.) Second, we observe that $\tau_b^*$ is almost independent of $V$. In contrast, $\tau_b^*$ depends on the distribution of the small bubbles in the fluid. Indeed, $\tau_b^*$ increases slightly when the fluid column is stirred each time the flow rate is set to a new value (Fig.~\ref{fig.taubstar}), which reveals the sensitivity of the phenomenon to the rheology of the fluid. 
 
The existence of a characteristic time $\tau_b^*$ associated with the lifespan of the bubbling regime is quite
intuitive but the mechanism leading to the formation of the channel is complex. A full description of the phenomenon,
which would deserve an extensive study, is beyond the scope of the present article. However, we propose a simple heuristic argument (Appendix B) which suggests to describe the experimental data by the following fitting function
(Fig.~\ref{fig.taubstar}):
\begin{equation}
\tau_b^* \simeq -\tau_0 \ln{\Biggl[Q_c\Biggl(\frac{1}{Q^*}-\frac{1}{Q}\Biggr)\Biggr]}.
\label{temps}
\end{equation} 
The key ingredients are a decrease of the flow threshold (yield stress) due to the shear imposed by the rising bubbles and the thixotropy of the fluid charged with bubbles. Interestingly, the rheological measurements reported in the appendix A point out that the small bubbles enhance the thixotropy and thus play here an important role. From the interpolation of the experimental data, we obtain $\tau_0 \simeq 65$~s, $Q_c = 0.35$~cm$^3$/s and $Q^* \simeq 0.29$~cm$^3$/s (added-water content $10\%$, $V=530$~cm$^3$ and $V=800$~cm$^3$, $h_0=70$~mm).
Interestingly, requiring $\tau_b^*>0$ in the equation (\ref{temps}), we obtain a maximum flow rate $Q^+ \equiv Q_c Q^*/(Q_c-Q^*) \simeq 1.7$~cm$^3$/s,
which roughly corresponds to the upper limit in figure~\ref{fig.pourcent}. In order to define this  upper limit properly, one must consider instead the flow rate for $\tau_b^*$ to equal the typical time of a bubble emission. We estimate, in this case, that the channel forms immediately, after the emission of a single bubble, for $Q>1.5$~cm$^3$/s. 

Thus, even if this rough model must be considered with caution, the proposed mechanism accounts for the existence of a characteristic lifespan $\tau_b^*$ of the bubble regime in a finite range $]Q^*,Q^+]$ of the imposed flow rate,  $\tau_b^*$ decreasing when the flow rate $Q$ is increased.

\subsection{Winding flue regime}

\begin{figure}[!t]
\begin{center}
\includegraphics[width=0.9\columnwidth ]{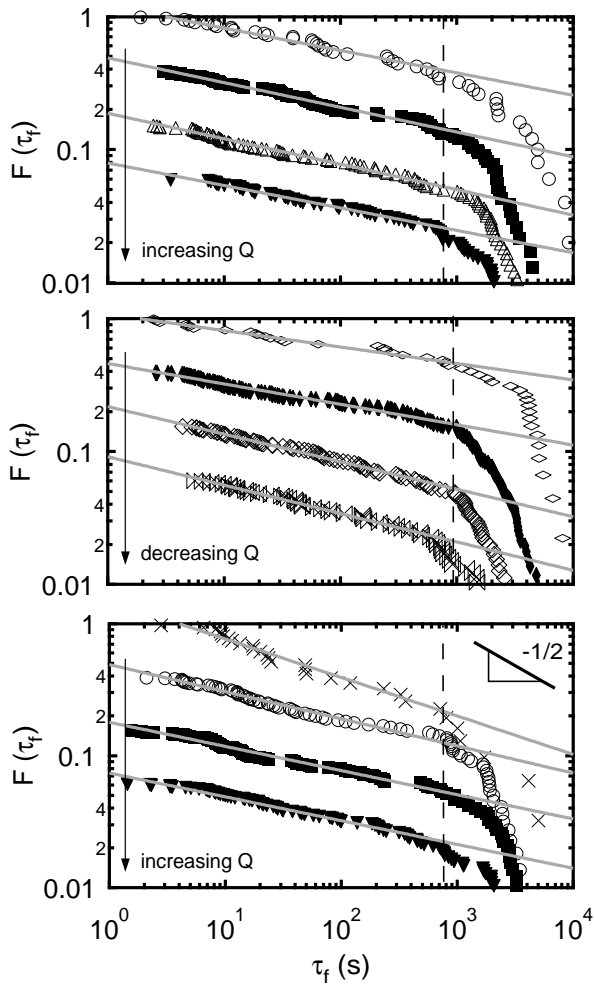}
\end{center}
\caption{
   \small {\bf Cumulative probability distribution $F_f(\tau_f)$ -}
   The cumulative probability distribution $F_f(\tau_f)$ of the time spent in the flue regime exhibits a power-law behavior, up to a large-time cutoff $\tau_2$. Curves have been vertically shifted for clarity, and the slope $1/2$ characterizing standard {\it on-off} intermittency is shown for comparison. Note that the data correspond to an acquisition time of $4$ days, which is significantly larger than $\tau_2$, so that we expect the system to have reached a stationary state \cite{Bertin} and the cut-off not to be due to the finite experimental time [(Symbol, $Q$ cm$^3$/s). Top: ($\circ$, 0.521); 
		  ($\blacksquare$, 0.695); ($\bigtriangleup $, 0.869); 
		  ($\blacktriangledown$,~1.04). Middle: ($\diamond$,1.22); ($\blacklozenge$,~0.782); 
		  ($\diamond$, 0.608); ($\Join$, 0.434). Bottom: ($\times $,~0.347); ($\circ$,~0.521);
		  ($\blacksquare$,~0.695); ($\blacktriangledown$,~1.04). (Added-water content $10\%$, $V=~530$~cm$^3$, $h_0=70$~mm)].}
\label{fig.distriCO}
\end{figure}

\begin{figure}[t]
\begin{center}
\includegraphics[width=0.9\columnwidth ]{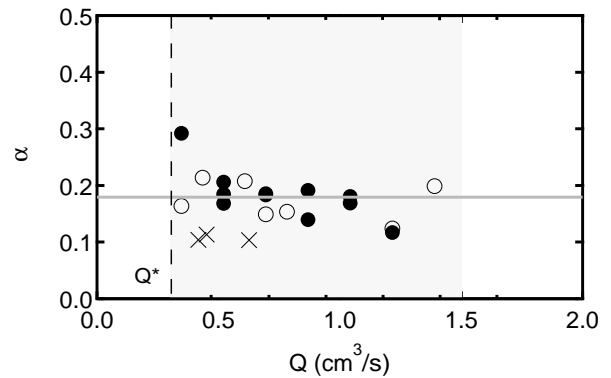}
\end{center}
\caption{
   \small {\bf Exponent $\alpha$ vs. flow rate Q - }
 The values of $\alpha$ obtained for increasing and decreasing flow rates (respectively closed and open symbols) are almost the same and significantly smaller than 1/2. Stirred samples (crosses) lead to even smaller values of $\alpha$
  (added-water content $10\%$, $V=530$~$cm^3$ and $V=800$~cm$^3$, $h_0=70$~mm).}
\label{fig.alpha}
\end{figure}

From the temporal sequence $\{\tau_f^{(n)}\}$, we observe that, for $Q \in~]Q^*,Q^+]$, the cumulative distribution $F_f(\tau_f)$ of the winding-flue lifespan $\tau_f$ presents a power-law behavior, $F_f(\tau_f) \propto \tau_f^{-\alpha}$ for $\tau_f \in [\tau_1,\tau_2]$ with $\tau_1 \sim 10$~s and $\tau_2 \sim 1000$~s (Fig.~\ref{fig.distriCO}) and a rapid decay beyond the cut-off time $\tau_2$.
We report in the figure ~\ref{fig.alpha} the experimental values of $\alpha$ obtained for flow rates in the gray region defined in the figure~\ref{fig.pourcent}. We estimate that $\alpha = 0.18\pm0.02$, this latter value being independent of $Q$ and $V$. However, $\alpha$ depends on the spatial distribution of the small bubbles which alter the rheological properties of the fluid (Appendix A). This conclusion is drawn by using a slightly different protocol: in an experimental series, we stirred the solution after each increment in the flow rate and estimated an even smaller value $\alpha \simeq 0.1$ in this latter case (Fig.~\ref{fig.alpha}). We thus observe that $\alpha$ is always significantly lower than the usual value $\alpha=0.5$, often reported in experimental observations \cite{Hammer94,Rodelsperger95,Feng98,John99,Vella03,Dickson97,Brokmann03,Zumofen04} and also expected from some of the theoretical intermittency scenarios, like the {\it on-off} intermittency \cite{Fujisaka1985,Platt1993,Heagy1994}.
Note that such a small value of the exponent ($\alpha<1$) imposes a cut-off at long times, $\tau_2$, which is necessary for the average $\langle \tau_f\rangle$ to exist. Experimentally, the data correspond to an acquisition time of $4$ days, which is significantly larger than $\tau_2$, so that we expect the system to have reached a stationary state \cite{Bertin} and the cut-off not to be due to the finite experimental time.
In some sense, $\tau_2$ corresponds to the largest memory time scale during which the stresses in the fluid at rest remain correlated, which is compatible with the viscoelastic time of the fluid, that is to say $\tau_2 \sim \tau_c$, thus of the order of 10$^3$~s (Appendix A). 

Hence, the flue lifespan does not exhibit any characteristic timescale in a temporal window between
$\tau_1$ and $\tau_2$ in which a power-law distribution with a small exponent is observed. 
These observations, which lie at the core of the present article, are thoroughly discussed in the section \ref{discussion}.

\section{Discussion}
\label{discussion}

As explained above, our main result is the experimental observation
that the probability density function
of flue lifespans displays a power-law behavior over a broad time window,
with a non-standard exponent $1+\alpha \sim 1.2$,
which slightly depends on the rheological properties of the fluid.
This result is of particular interest, since in standard intermittency
scenarios, the exponent takes a universal value,
independent of the details of the dynamics. For instance, in the
{\it on-off} intermittency scenario, the exponent $1+\alpha=3/2$ is directly
related to the first return time probability of a random walk
(or a ``chaotic walk'' in the deterministic case) \cite{Heagy1994}.
In purely deterministic scenarios, like type-I, type-II and type-III
intermittency scenarios \cite{Pomeau1980,Schuster1989},
the exponents are related to the
classification of bifurcations in dynamical system theory
\cite{Pomeau1980,Schuster1989,Heagy1994}.
The corresponding values of the exponent $1+\alpha$ in these scenarios
($1/2$ for type-I intermittency, $2$ for type-II and $3/2$ for type-III
\cite{Schuster1989,Heagy1994})
all differ significantly from our experimental values.

In the present experiment, due to both its non-standard value
and its observed dependence on the experimental conditions,
$\alpha$ is likely to be related to some rheological
or physical properties of the system, rather than to some
generic theoretical property.
It is thus interesting to relate the measured value of
the exponent $\alpha$ to
some properties of the gel. A full analysis of this issue
is well beyond the scope of the present article, but
in order to get some first insights, we propose to describe the data
with a simple stochastic model with two states, namely 
{\it bubble} or {\it flue}.
In such a framework, we write that, after having spent a time $\tau_b$ in the {\it bubble} regime, the system has a probability per unit time (or transition rate) $r_b(\tau_b)$ to switch to the {\it flue} regime.
Similarly, we assume that 
after a time $\tau_f$ in the {\it flue} regime,
the transition rate to the {\it bubble} regime is $r_f(\tau_f)$.
In the {\it bubble} regime, the exponential distribution, $F_b(\tau_b)$,
with the characteristic time $\tau_b^*$ means that the transition rate $r_b$
is constant, independent of $\tau_b$, and equal to $r_b=1/\tau_b^*$.
In contrast, in the {\it winding-flue} regime, the observed power-law
distribution $F_f(\tau_f)$ implies that the transition rate $r_f(\tau_f)$
is not a constant, and thus that a more complex dynamics is at play,
even in this phenomenological description. From the relation
\be \label{eq-diff-F}
\frac{dF_f}{d\tau_f} = -r_f(\tau_f) \, F_f(\tau_f),
\ee
a power-law behavior $F_f(\tau_f) \propto \tau_f^{-\alpha}$ is found to
correspond to $r_f(\tau_f)=\alpha/\tau_f$.
Still, the power-law behavior is observed only in a window
$\tau_1 \lesssim \tau _f \lesssim \tau_2$, which suggests that,
as a first approximation, $r_f$ could be taken as a constant outside
this window. A simple parameterization of $r_f(\tau_f)$ leading to such
a behavior is given by
\be
r_f(\tau_f) = \frac{1}{\tau_0 (1+\tau_f/\tau_1)} + \frac{1}{\tau_2}
\ee
provided $\alpha=\tau_1/\tau_0$ and $\tau_1 \ll \tau_2$.
The resulting distribution
\be
F_f(\tau_f) = \frac{e^{-\tau_f/\tau_2}}{(1+\tau_f / \tau_1)^{\alpha}}
\ee
behaves as a power law for $\tau$ larger than $\tau_1$, with an exponential
cut-off at $\tau_2$.
In the opposite limit $\tau_f \ll \tau_1$, $F_f(\tau_f)$ can be expanded as
\be
F_f(\tau_f) \simeq 1-\tau_f/\tau_0.
\ee
This linear decay of $F_f(\tau_f)$ at short time can thus be used to evaluate
the time scale $\tau_0$. In addition, the time scale $\tau_1$ can be
estimated from the crossover between the linear and power-law behavior.
From the experimental data, we assess $\tau_0 \sim 30$-$40~$s and
$\tau_1 \sim 5$-$8~$s. These values lead to a
ratio $\tau_1/\tau_0$ in the range $0.12$-$0.27$,
compatible with the experimental value of $\alpha$.
We can thus consistently interpret the exponent $\alpha$ as the ratio
of two characteristic time-scales: the time $\tau_1$ beyond which the winding
flue starts to age (roughly the time for the flue to reach its stationnary shape)
and the time $\tau_0$ typically needed for the flue to collapse in the
early stationary regime. Finally, note that the time $\tau_0$ is of the same
order of magnitude as the time needed for the yield stress of the fluid to relax to 
its asymptotic value $\sigma_c$ after the shear induced by the rise of a single 
bubble (see appendix \ref{estimate}).

\section{Conclusion and Outlooks}

We have shown that the flow of a Newtonian fluid into a non-Newtonian
one can exhibit an intermittent activity, namely between a {\it bubbling}
and a {\it winding flue} regime.
The distribution of the {\it off}-state lifespan displays a power-law behavior
with an exponent about $1.2$, which significantly differs
from the standard exponent $3/2$.
A simple stochastic two-states model suggests that this non-standard exponent
could be directly related to the ratio of two timescales characterizing
the experimental system.

Our results also contain relevant conclusions in a geophysical context, even if we do not claim at any direct interpretation. We propose that the volcanic intermittent activity observed on the field, which has been hitherto explained in terms of constriction of the magma chamber \cite{Jaupart} or by changes in gas flux and/or magma flow rate at depth \cite{Ripepe,Woods94}, could be (partly) the results of the non-Newtonian properties of magma. Indeed, our simple laboratory experiment demonstrates that, even if a constant gas flow is imposed, one can observe an intermittent activity on top of the fluid column.

A more detailed understanding of this non-standard intermittency phenomenon would certainly be desirable, both from  practical and fundamental viewpoints. An extensive study of the dependence of the exponent $\alpha$ on the physical properties of the fluid (rheological properties, surface tension \cite{Tzounakos}, \ldots) is in progress. Moreover, we are also currently investigating a two-dimensional experiment, similar to the present three-dimensional one, in which one can monitor the evolution and the collapse of the flue. We aim at relating the overall dynamics of the system to the local dynamics. 

Finally, we proposed that the minimum ingredients necessary for a channel to form and stabilize, and thus for the intermittency to be observed, are a shear-thinning and thixotropic fluid. However, a detailed study of the mechanisms at stake in the flue formation and collapse remains to be performed. 



\begin{acknowledgments}

The authors appreciated remarks from J.-F. Palierne and numerous enlightening discussions with S. Vergniolle. T.D. and V.V. have also greatly benefited from the IAVCEI General Assembly which took place in Reykjavik in late August 2008.

\end{acknowledgments}

\appendix{}

\section{Fluid characterization}

\subsection{Rheology of the fluid}
\label{rheo}

\begin{figure}[h!]
\begin{center}
\includegraphics[width=0.9\columnwidth ]{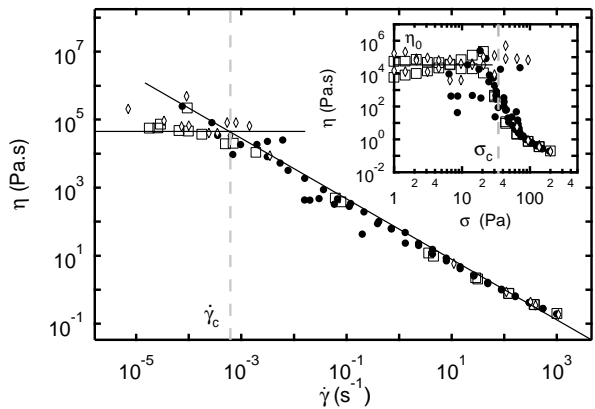}
\end{center}
\caption{\small{{\bf Fluid viscosity $\eta $ vs shear-rate $\dot \gamma$ -} The fluid is shear-thinning. The vertical gray line indicates the characteristic shear-rate $\dot \gamma_c$. {Inset: Fluid viscosity $\eta $ versus shear stress $\sigma$ -} The measurements clearly point out the existence of a flow threshold $\sigma _c \simeq 35$~Pa [Open symbols correspond to stress-controlled experiments whereas solid symbols correspond to shear-controlled experiments. Gap $e = 750$~$\mu$m : ($\square$, $\Delta t = 25$ s); ($\diamondsuit$, $\Delta t = 1$ s); ($\bullet$, $\Delta t = 10$~s)].
}}
\label{fig.rheology}
\end{figure}

The rheological properties of the solution, free of bubbles, have been measured using a rheometer (Bohlin Instruments, C-VOR 150) equipped with parallel plates (PP-60, gap $e = 750$ and 1000 $\mu$m). Sand paper was glued to the plates in order to prevent any sliding of the fluid at the walls; At 25$^{\circ}$C, we performed both steady-shear and steady-stress experiments with an equilibration time $\Delta t$ ranging from 1 to 25~s (Fig.~\ref{fig.rheology}). The solution is shear-thinning and we estimated the associated characteristic time $\tau_c = 1/\dot \gamma _c \simeq 1700$~s, compatible with the ratio of the viscosity at low shear-rate $\eta_0 \simeq 6 \times 10^4$~Pa.s to the estimated flow threshold $\sigma _c \simeq 35$~Pa (Fig.~\ref{fig.rheology}).

\subsection{Bubbles size and density}

During the experiments, the small bubbles that remain trapped in the fluid column, due to 
the yield stress, alter the rheological properties of the fluid.We report here measurements 
of the size and density of the small bubbles along the fluid column when the system has reached 
a steady state.

During the first cycle of $Q$, the state of the column evolves as it fills up with small bubbles which get trapped. However, after having switched $Q$ twice to $Q^*$ during the increase and the decrease of the flow rate, the aspect of the column remains qualitatively the same, and one can assume that the vertical bubble gradient has reached a steady state (We give several indications of such an assertion in Sect.~\ref{Experimental}.) In order to characterize the vertical bubble gradient which takes place along the column height in the steady-state regime, we perform the following experiment: from a $15$~cm high column which has experienced 5 flow rate cycles, we carefully extract 7~slices, 2~cm thick, corresponding to different altitude $z$; For each slice, we measure the size of a thousand bubbles using a macroscope ({\it Leitz, Laborlux}) and assess the fluid density. The corresponding rheological properties shall be reported in the section \ref{rheobulle}.
\begin{figure}[h]
\begin{center}
\includegraphics[width=0.9\columnwidth ]{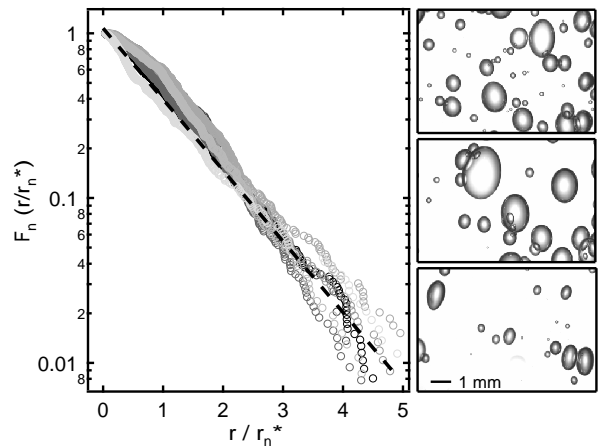}
\end{center}
\caption{\small{{\bf Bubbles size and density -} Left: Cumulative probability distribution $F_n$ of the bubble size $r$ in the slice $n$ (The grayscales are uniformely distributed  from light gray at the bottom to black at the top.) The distribution $F_n$ is exponential (linear fit, dotted black-line.) We denote $r_n^*$ the associated characteristic radius. {Right: } The pictures, from bottom to top, correspond respectively to the slices, $z \in [0.5,2.5]$~cm, $z\in[8.5,10.5]$~cm and $z \in [12.5,14.5]$~cm, respectively.
}}
\label{fig.Bulle}
\end{figure}

\begin{figure}[h] 
\begin{center}
\includegraphics[width=0.9\columnwidth ]{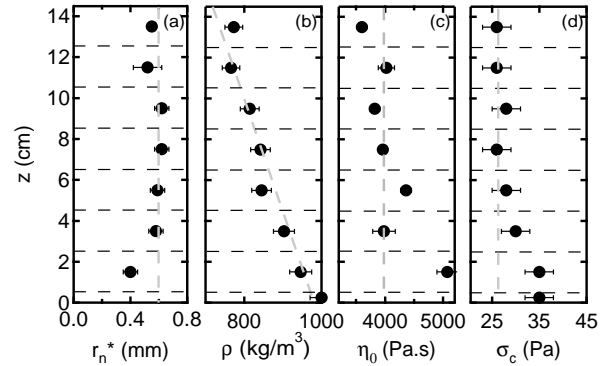}
\end{center}
\caption{\small{\bf Fluid column properties -} We report (a) the mean bubble size $r_n^*$, (b) the density $\rho$, (c) the viscosity $\eta_0$ in the limit of low shear-rate $\dot \gamma$, and (d) the fluid-threshold $\sigma_c$ along the column height $z$. Each point corresponds to an average over one slice (thickness 2~cm, horizontal dashedlines). 
}
\label{fig.caracterisation}
\end{figure}

Let us denote as $p_n(r)$ the probability distribution of the bubble size $r$ in the slice $n$. In each slice, the cumulative probability distribution of the bubble size $r$, defined as $F_n(r) \equiv \int_{r}^{\infty} p_n(r')\,dr'$ is exponential (Fig.~\ref{fig.Bulle}). Thus, the bubble size, in each slice, is exponentially distributed. The corresponding characteristic radius, $r_n^* = (0.58 \pm 0.04)$~mm, is almost constant over the column height $h$ (Fig.~\ref{fig.caracterisation}a). In order to assess the associated bubble concentration, we measured the density of material in each slice (Fig.~\ref{fig.caracterisation}b). The fluid density decreases roughly linearly from the density of the fluid free of bubbles at the bottom to $\rho \simeq 750$~kg/m$^3$ at the free surface, corresponding to a maximum gas fraction of about 25\%. The measurements confirm the visual observation of a bubble gradient inside the fluid. We estimate that the bubble concentation, which equals 0 at the bottom, is about 300 bubbles per cm$^{3}$ at the top.

\subsection{Rheological properties in presence of bubbles}
\label{rheobulle}

\begin{figure}[t]
\begin{center}
\includegraphics[width=0.9\columnwidth ]{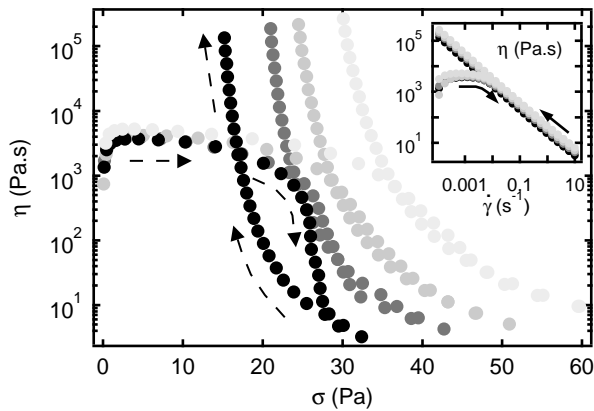}
\end{center}
\caption{\small{{\bf Viscosity $\eta$ vs shear stress $\sigma$ - } The rheological properties depend on the altitude $z$ because of the variation of the density of the bubbles trapped in the fluid along the column height. Moreover, we observe that the data obtained for increasing and decreasing shear rate (arrows) do not superpose: the presence of the bubbles leads to a significant memory effect. {Inset:} Fluid viscosity $\eta $ vs shear-rate $\dot \gamma$ - The measurements clearly point out the memory effect (The symbols, from light gray to black, correspond respectively to the slices, $z \in [0.5,2.5]$~cm, $z\in[3.5,5.5]$~cm, $z\in[8.5,10.5]$~cm and $z \in [12.5,14.5]$~cm, where $z$ denotes the altitude. Imposed shear-rate, equilibration time $\Delta t = 15$~s).
}}
\label{fig.rheobulle}
\end{figure}

Using the same experimental protocol as in the section \ref{rheo},
we determined, for each slice, the rheological properties of the fluid, charged with bubbles. We observe that the bubbles lead to the appearence of a significant memory effect (Fig.~\ref{fig.rheobulle}): the shear stress does not reach the steady-state value instantaneously but after a significant characteristic time, $\tau_0$, which we estimate from the experimental data to be about 1 min for the largest bubble density (at the top of the column). In contrast, we observe that the shear-thinning properties of the fluid are not qualitatively altered (Fig.~\ref{fig.rheobulle}, inset).
Note finally that the yield stress $\sigma_c \simeq 25$~Pa is roughly constant along the column (Fig.~\ref{fig.caracterisation}d), and smaller that the estimated value for the free bubble gel solution ($\sigma_c \simeq 35$~Pa).

\section{Estimate of $\tau_b^*$}
\label{estimate}

In order to account, at least at a rough estimate, for the observation of a characteristic time of the bubbling regime, we propose that a channel can form and remain stable if the shear-stress imposed by the inner gas flow at the channel walls exactly balances the yield stress. Initially, the yield stress is large and the gas emission at the nozzle necessitates a large overpressure which leads to the formation of a bubble once the fluid starts flowing. The rise of successive bubbles leads to a decrease in the yield stress which allows the emission of gas at a small overpressure and, thus, the formation of a thin channel in the fluid. Moreover, once the channel has formed, there is no flow of the fluid at the walls and the yield stress increases back to its maximum value, which tends to stabilize the structure.

On the one hand, between two successive bubbles, the fluid is at rest and the yield stress, $\sigma_y$, relaxes toward its asymptotic value $\sigma_c$ over the characteristic time $\tau_0$ according to
\begin{equation}
\tau_0 \dot\sigma_y + \sigma_y = \sigma_c. 
\label{relax}
\end{equation}
where $\tau_0$ is an intrinsic property of the fluid.
On the other hand, due to the large shear stress imposed by the rise of the bubble, $\sigma_y$ tends to vanish according to
\begin{equation}
\tau_b \dot\sigma_y + \sigma_y = 0. 
\label{destr}
\end{equation}
where $\tau_b$ is here not an intrinsic property of the fluid. Indeed, $\tau_b$ characterizes the destructuration of the fluid under stress and, for instance, might depend on the rising-bubble size. Hence, $\tau_b \le \tau_0$ \cite{Barnes97}.

Considering the typical time, $\tau \equiv V_b/Q$, between two successive bubbles ($V_b$ denotes the volume of the bubble) and the duration of the shear by a rising bubble, $t_b$, according to equations (\ref{destr}) and (\ref{relax}), one can estimate that, after the rise of $n$ successive bubbles:
\begin{equation}
\sigma_y \simeq \sigma_c - \sigma_c \frac{t_b}{\tau} \frac{\tau_0}{\tau_b} \Bigl[1 - \exp{\Bigl(-n \frac{\tau}{\tau_0}\Bigr)}\Bigr]
\label{stress}
\end{equation} 
where we assumed that $t_b \ll \tau \ll \tau_b$ and only kept the exponential dependence on $n$.

When the yield stress has reached a small enough value, the gas emission can occur in a gentle manner, 
a thin gas finger growing from the nozzle. We can estimate that the growth of the finger will lead to a stable channel provided that the shear stress applied to the wall by the inner gas flow does not exceed the yield stress. Assuming a Poiseuille flow inside, one can estimate the shear stress at the wall $\sigma_p = 4 \eta_a Q /(\pi  r_c^3)$ where $\eta_a = 1.6 \times 10^{-5}$~Pa.s denotes the dynamic viscosity of air and $r_c$ the radius of the channel. In this framework, the condition $\sigma_p = \sigma_y$ leads to:  
\begin{equation}
\tau_b^* = - \tau_0 \ln{\Bigl[1 - \frac{\tau}{t_b} \frac{\tau_b}{\tau_0} \Bigl(1 - \frac{\sigma_p}{\sigma_c}\Bigr) \Bigr]} 
\label{duree}
\end{equation}
where we defined $\tau_b^* \equiv n \tau$. This latter equation can be rewritten :
\begin{equation}
\tau_b^* \simeq -\tau_0 \ln{\Biggl[Q_c\Biggl(\frac{1}{Q^*}-\frac{1}{Q}\Biggr)\Biggr]}
\label{temps-app}
\end{equation} 
where, by definition, 
\begin{equation}
Q_c \equiv \frac{\tau_b}{\tau_0} \frac{V_b}{t_b} \hspace{0.5cm} {\rm and} \hspace{0.5cm} Q^* \equiv \frac{Q_c}{1+Q_c/Q^+}
\end{equation}
with $Q^+\equiv \frac{\pi r_c^3 \sigma_c}{4 \eta_a}$. 
Note that $\tau_b^*$ ($>0$) is thus predicted to be defined in a finite range $Q \in~]Q^*,Q^+]$ of the flow rate and, in agreement with the experimental observations, not to depend on the height of the fluid column.

From the interpolation of the experimental data (Fig.~\ref{fig.taubstar}), $Q_c \simeq 0.35$~cm$^3$/s, $Q^* \simeq 0.29$~cm$^3$/s and $\tau_0 \simeq 65$~s. First, we observe that $\tau_0$ is found to be in reasonable agreement with the estimation obtained in the rheology experiment. Second, from the time difference between successive bubbles in the pressure signal (Fig.~\ref{fig.2}), we obtain the typical bubble-volume $V_b \simeq 2$~cm$^3$ and, accordingly, the typical radius $r_b \simeq 0.8$~cm. We estimate the rise time of the bubble along the whole column height (8 cm) to be about 1~s and, thus, $t_b \simeq 0.2$~s (about $2 r_b / v_b$, the diameter divided by the rise velocity, $v_b \simeq 8$~cm/s). With this latter value of $t_b$ and  the experimental $Q_c$, we obtain $\tau_b \simeq 2.3$~s, which is a reasonnable value if one considers that the shear rate imposed by the rising bubble is large, of about $1/t_b = 5$~s$^{-1}$ (Note, in addition, that the imposed stress is of about $\rho g r_b \sim 80$~Pa where $\rho$ denotes the density of the fluid and $g$ the acceleration due to the gravity.) From $Q^+ \simeq 1.7$~cm$^3$/s and $\sigma_c \simeq 25$~Pa (Fig.~\ref{fig.caracterisation}), one can estimate the channel radius $r_c \simeq 0.1$~mm, which is also in agreement with the experimental observation of a very thin channel, a fraction of millimeter in diameter. At last, it would be interesting to understand the mechanism that selects the channel radius (competition between the elasticity and the surface tension?) but the problem is out of the scope of this appendix.

From the proposed mechanism, one can also understand why the small bubbles quickly invade the system when $Q$ is about $Q^*$
(i.e. $Q_b \simeq Q^*$, Sec.~\ref{po}). Indeed, on the one hand, in the bubbling regime, the pinch-off of the bubble emitted at the nozzle \cite{Thoroddsen}, the coalescence of two bubbles in the bulk (\cite{Zhang}, Sec.~\ref{Experimental}) and the bursting of a bubble at the free surface \cite{Divoux} are likely to leave a small satellite (bubble) \cite{remark1}. We can thus guess that each rising bubble leaves one or two small bubbles. On the other hand, in the flue regime, the number of small bubbles left by the channel collapse is about the column height divided by the wavelength that develops (Rayleigh instability \cite{Eggers97}). In our experimental conditions, we estimate from direct visual observation that the wavelength is about 1 cm (much larger than the channel diameter) and, thus, that each collapse of a channel produces about 10 small bubbles. The channel collapse is thus much more efficient in the production of small bubbles. By definition of $Q^*$, at the transition, the channel is marginaly stable and then frequently collapses. As a conclusion, $Q_b \simeq Q^*$.


\begin{thebibliography}{}

\bibitem{Hammer94}
P.~W. Hammer, N. Platt, S.~M. Hammel, J.~F. Heagy, and B. D. Lee ,
Phys. Rev. Lett. \textbf{73}, 1095 (1994).

\bibitem{Rodelsperger95}
F. R\"odelsperger, A. Cenys, and H. Benner, Phys. Rev. Lett. \textbf{75},
2594 (1995).

\bibitem{Feng98}
D.~L. Feng {\it et al.}, Phys. Rev. E
\textbf{58}, 3678 (1998).

\bibitem{John99}
T. John, R. Stannarius, and U. Behn, Phys. Rev. Lett. \textbf{83},
749 (1999).

\bibitem{Vella03}
A. Vella {\it et al.}, Phys. Rev. E
\textbf{67}, 051704 (2003).

\bibitem{Dickson97}
R.~M. Dickson \textsl{et al.}, Nature (London) \textbf{388}, 355 (1997).

\bibitem{Brokmann03}
X. Brokmann, J.-P. Hermier, \textsl{et al.}, Phys. Rev. Lett. \textbf{90},
120601 (2003).

\bibitem{Zumofen04}
G. Zumofen, J. Hohlbein and C.~G. Hubner, Phys. Rev. Lett. \textbf{93},
260601 (2004).
 
\bibitem{Lo2002}
C.-C. Lo, L. A. Nunes Amaral, S. Havlin, {\it et al.} Europhys. Lett.
\textbf{57}, 625 (2002).

\bibitem{Fujisaka1985}
H. Fujisaka and T. Yamada, Prog. Theor. Phys. \textbf{74}, 918 (1985).

\bibitem{Platt1993}
N. Platt, E.~A. Spiegel, and C. Tresser, Phys. Rev. Lett. \textbf{70},
279 (1993).

\bibitem{Heagy1994}
J.~F. Heagy, N. Platt, and S.~M. Hammel, Phys. Rev. E \textbf{49},
1140 (1994).

\bibitem{Pomeau1980}
Y. Pomeau and P. Manneville, Commun. Math. Phys. \textbf{74}, 189 (1980).

\bibitem{Schuster1989}
H.~G. Schuster, \textit{Deterministic Chaos}, 2nd Edition
(VCH, New York, 1989).

\bibitem{Zoueshtiagh}
F. Zoueshtiagh and A. Merlen, Phys. Rev. E \textbf{75}, 056313 (2007). 

\bibitem{Gostiaux}
L. Gostiaux, H. Gayvallet and J.-C. G\'eminard, Granular Matter \textbf{4},
39 (2002).

\bibitem{Varas}
G. Varas, V. Vidal and J.-C. G\'eminard,
Phys. Rev. E {\bf 79}, 021301 (2009).

\bibitem{Ripepe}
M. Ripepe, S. Ciliberto, M. Della Schiava, J. Geophys. Res.
{\bf 106}, 8713 (2001); M. Ripepe, A.J.L. Harris, R. Carniel, J. Volc.
Geotherm. Res. {\bf 118}, 285 (2002).

\bibitem{Gonnermann}
H.~M. Gonnermann and M. Manga, Annu. Rev. Fluid Mech.
\textbf{39}, 321 (2007).

\bibitem{Webb90}
S.~L. Webb and D.B. Dingwell, J. Geophys. Res. {\bf 95}, 695 (1990).

\bibitem{Lavallee2007}
Y. Lavall\'ee, K.-U. Hess, B. Cordonnier, and B.~D. Dingwell, Geology, {\bf 35}, 843-846 (2007). 

\bibitem{Parfitt}
E.~A. Parfitt, J. Volc. Geotherm. Res. \textbf{134}, 77 (2004).

\bibitem{Lavallee2008}
Y. Lavall\'ee, P.~G. Meredit, B.D. Dingwell {\it et al.} Nature, {\bf 453}, 507 (2008). 

\bibitem{Caricchi08}
L. Caricchi, D. Giordano, L. Burlini, P. Ulmer and C. Romano, Chem. Geology, {\bf 256}, 158 (2008).

\bibitem{Jaupart}
C. Jaupart and S. Vergniolle, Nature \textbf{331}, 58 (1988); J. Fluid. Mech. \textbf{203}, 347 (1989).

\bibitem{Okumura2006}
S. Okumura, M. Nakamura and A. Tsuchiyama, Geophys. Res. Lett. {\bf 33}, L20316 (2006).

\bibitem{Okumura2008}
S. Okumura, M. Nakamura, A. Tsuchiyama, T. Nakano and K. Uesugi, J. Geophys. Res. {\bf 113}, B07208 (2008).

\bibitem{Tufaile} 
A. Tufaile and J. Sartorelli, Phys. Rev. E {\bf 66}, 056204 (2002).

\bibitem{Kliakhandler}
I.~L. Kliakhandler, Phys. Fluids \textbf{14}, 10 (2002).

\bibitem{Pinkerton95}
H. Pinkerton and G. Norton, J. Volc. Geotherm. Res. {\bf 68}, 307-323 (1995). 

\bibitem{Divoux}
T. Divoux, V. Vidal, F. Melo and J.-C. G\'eminard, Phys. Rev. E {\bf 77} 056310 (2008).

\bibitem{Daugan}
S. Daugan, L. Talini, B. Herzhaft and C.Allain, Eur. Phys. J. E {\bf 7}, 73 (2002);
Eur. Phys. J. E. {\bf 9}, 55 (2002).

\bibitem{Berg}
B.~A. Berg and R.C. Harris, Comput. Phys. Commun. {\bf 179}, 443–448 (2008).

\bibitem{Bertin}
E. Bertin and F. Bardou, Am. J. Phys. \textbf{76}, 630 (2008).

\bibitem{Woods94}
 A.W. Woods and T. Koyaguchi, Nature {\bf 370}, 641 (1994).
 
\bibitem{Tzounakos}
A. Tzounakos, D.~G. Karamanev, A. Margaritis and M.~A. Bergougnou, Ind. Eng. Chem. Res. {\bf 43}, 5790 (2004).

\bibitem{Barnes97}
H.A. Barnes, J. Non-Newtonian Fluid Mech., {\bf 70}  1-33 (1997).

\bibitem{Thoroddsen}
S.~T. Thoroddsen, T.~G. Etoh and K. Takehara, Phys. Fluids {\bf 19}, 042101 (2007).

\bibitem{Zhang} 
F.~H. Zhang and S.~T. Thoroddsen, Phys. Fluids {\bf 20}, 022104 (2008).

\bibitem{remark1} {Please note that the formation of satellite by bubble pinch-off and bubble coalescence has been studied in the case of a Newtonian fluid and is still an active field of research \cite{Thoroddsen,Zhang, Burton}; This topic still remains to be fully characterized in the case of non-Newtonian fluids.}








\bibitem{Burton}
J.~C. Burton and P. Taborek, Phys. Rev. Lett. {\bf 101}, 214502 (2008).


\bibitem{Eggers97} J. Eggers, Rev. Mod. Phys. {\bf 69}, 865 (1997).






\end{thebibliography}
\end{document}